\documentclass{appolb}
\usepackage[utf8]{inputenc}
\usepackage{float}
\usepackage{sidecap}
\usepackage{amsmath,amssymb,amsfonts,bm}
\usepackage{graphicx}
\usepackage{multirow}
\usepackage[usenames,dvipsnames]{color}
\usepackage[
colorlinks=true,
linkcolor=blue,
breaklinks=true,
urlcolor=blue,
citecolor=blue]{hyperref}
\usepackage{orcidlink}
\usepackage{graphicx,color,rotating}
\usepackage{dcolumn}
\usepackage{bm}
\newcommand{\er}{$\pm$}
\newcommand{\bc}           {\begin{center}}
\newcommand{\ec}           {\end{center}}
\newcommand{\bq}           {\begin{eqnarray}}
\newcommand{\eq}           {\end{eqnarray}}
\newcommand{\be}           {\begin{equation}}
\newcommand{\ee}           {\end{equation}}
\newcommand{\bi}           {\begin{itemize}}
\newcommand{\ei}           {\end{itemize}}
\definecolor{darkred}{rgb}{0.7,0,0}
\definecolor{darkgreen}{rgb}{0,0.4,0}
\definecolor{darkblue}{rgb}{0,0,0.4}
\definecolor{darkbrown}{rgb}{0.5,0,0}
\definecolor{darkcyan}{cmyk}{1,0.3,0.3,0.3}
\definecolor{midgreen}{rgb}{0,0.6,0}
\newcommand {\rd}{\color{red}}
\newcommand {\bl}{\color{blue}}
\newcommand {\gr}{\color{midgreen}}
\newcommand {\br}{\color{darkbrown}}

\begin{document}
\title{Dyons, Instantons, Baryons and AdS/QCD}%

\author{Eberhard Klempt
\address{ Helmholtz-Institute f\"ur Strahlen- und Kernphysik
der Rheinischen Friedrich-Wilhelms Universit\"at, Nussallee
14 - 16, 53115 Bonn, Germany\\
and\\
Thomas Jefferson National Accelerator Facility, Newport News, VA, USA}
\\[3mm]
} 

\maketitle
\begin{abstract}
Dmitri Diakonov has played a significant role in identifying the degrees of freedom
underlying hadron spectroscopy. His contributions are discussed with a view on recent
developments.
\end{abstract}
   
\section{Introduction}
In summer 2001, Mitya visited our institute in Bonn and gave a fascinating talk ~\cite{Diakonov:2002mb} about
instantons and baryon dynamics and explained how instantons break the chiral symmetry
of strong interactions~\cite{Diakonov:1987ty}, ideas Mitya had developed further with Maxim~\cite{Diakonov:1995qy}.
I was startled. At that time I just had found
a mass formula for baryon resonances based on instanton-induced interactions~\cite{Klempt:2002vp},
which reproduced the spectrum much better than quark models and with fewer parameters~\cite{Klempt:2010du}. 
I had the opportunity to explain to Mitya my new formula and some wild ideas I had about different time
scales for color and flavor exchange in baryons and the consequences for confinement, 
the $^3P_0$ model, the proton spin, for glueballs and hybrids~\cite{Klempt:2002cu}. Mitya listened with friendly
attention, and if he felt bored, he surely hided his feelings. 

Later, Mitya delevoped - jointly with Vitya - ideas to understand confinement by
gluon field configurations with asymptotic Coulomb-like chromo-electric and
-magnetic fields, called dyons~\cite{Diakonov:2004jn,Diakonov:2007nv}. Dyons are supposed to reveal
themselves as Abelian monopoles or as center vortices. The interactions of dyons
can be identified with instantons. In essence, in his view, mass generation, confinement, and
chiral symmetry breaking are related to a common origin.

\begin{figure}[pt]
    \centering
   \includegraphics[width=0.75\textwidth]{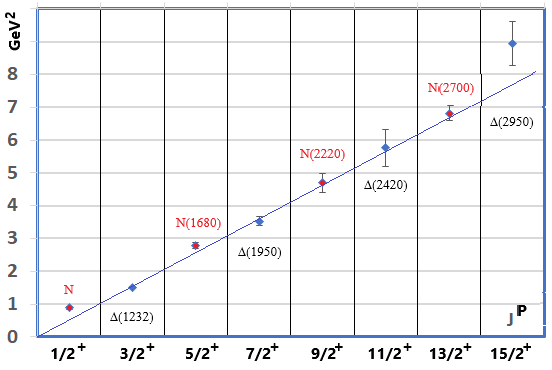}
    \caption{Squared masses (pole positions) of leading baryon resonances
versus total angular momentum $J$.}
    \label{regge}
\end{figure}

Figure~\ref{regge} supports this interpretation. It shows a common Regge trajectory 
for $N^*$ and $\Delta^*$ resonances, the leading $N^*$ resonances with $J=L+1/2$ and
$\Delta^*$ resonances with $J=L+3/2$. The masses, the Regge slope, and the hyperfine
structure splitting ($\Delta(1232)$ versus $N$, $\Delta(1950)$ versus $N(1680)$,
$\Delta(2420)$ versus $N(2220)$, $\Delta(2950)$ versus $N(2700)$) are consistently 
described, in agreement with Mitya's conjecture. But clearly, this is not yet the full
story: the fit is not perfect, and in particular the masses of nucleon and $N(1680)$ are not well
reproduced. Only the $\Delta^*$ values lie precisely on a straight line, coinciding 
with zero at $J=0$.

\section{Baryons in AdS/QCD}
In an ADS/QCD, approach developed by Brodsky, Teramond, Dosch, and Ehrlich~\cite{Brodsky:2014yha}, the spectrum of
baryon resonances is reproduced by one constant and two quantum numbers.  
The nucleon mass, the Regge slope, and the
mass splitting of resonances can all be described by just one parameter. 
This fact again points at a common origin of these three phenomena as Mitya, Vitya and
Maxim have advocated for.

The mass spectrum in AdS/QCD is given by a radial quantum number $n$ and
a quantum number $\nu$. 
\be
\label{ads}
M^2= 4\lambda(n+\nu+1)\,.
\ee
The calculated mass spectrum using eq.~(\ref{ads}) or, alternatively, the relations of Table~\ref{QM}
is compared to the data in Table~\ref{baryons}. $\nu$ needs to be related 
to the quantum numbers characterizing the different $N^*$ and $\Delta^*$ resonances
\cite{Brodsky:2014yha}, see Table~\ref{AdS-nu}. 

\begin{table}
	\caption{\label{baryons}Baryon resonances in the first and second excitation shell and their masses (all masses are given in MeV).
      Three- and four-star resonances are shown in bold font.  The assignments to the 20-plet are questionable.}
  	\renewcommand{\arraystretch}{1.5}
	\setlength{\tabcolsep}{5pt}
 \scriptsize
	\begin{center}
		\begin{tabular}{|c|c|c|c|c|c|l|r|l|r|} \hline
			\text{N}  & $L^P$ & SU(6) & $S$  & Baryon  & \multicolumn{1}{c|}{$M_{\rm c.o.g.}$}    & \multicolumn{2}{c|}{Instanton induced} & \multicolumn{2}{c|}{AdS/QCD}  \\ \hline\hline
 			
			$0$   & $0^+$ & $\mathbf{56}$ & 
			$\frac{1}{2}$ & \boldmath $N(940) \frac{1}{2}^+$       & \it 939  & 
            $\sqrt{M_\Delta^2 - \frac{\delta}{2}}$   & 939  & 
			$\sqrt{4\lambda}$         & 988 \\ \cline{4-10}
	
			      &       &               & 
			$\frac{3}{2}$ & \boldmath $\Delta(1232) \frac{3}{2}^+$ & \it 1210\er 1 &  
			$M_\Delta$  & 1210 & $\sqrt{6\lambda}$   & 1210 \\ \hline\hline

			$1$   & $1^-$ & $\mathbf{70}$ & $\frac{1}{2}$ & 
			\hspace{-1mm}\boldmath$\begin{array}{l} {\displaystyle N(1535)} \frac{1}{2}^- \\[-1mm] {\displaystyle N(1520)} \frac{3}{2}^-  \end{array}$    & \it 1503\er 4  & 
			$\sqrt{M_\Delta^2 + \omega - \frac{\delta}{4}}$  & 1572  & 
			$\sqrt{10\lambda}$   & 1562 \\ \cline{4-10}
			
                  &       &               &  $\frac{3}{2}$ & 
			\hspace{-1mm}\boldmath$\begin{array}{l} {\displaystyle N(1650)} \frac{1}{2}^- \\[-1mm] {\displaystyle N(1700)} \frac{3}{2}^- \\[-1mm] {\displaystyle N(1675)} \frac{5}{2}^- \end{array}$    & \it 1670\er 24  & 
			$\sqrt{M_\Delta^2 + \omega}$ & 1662 & 
			$\sqrt{12\lambda}$   & 1711 \\ \cline{4-10}
			
                  &       &               &  $\frac{1}{2}$ & 
            \hspace{-1mm}\boldmath$\begin{array}{l} {\displaystyle \Delta(1620)} \frac{1}{2}^- \\[-1mm] {\displaystyle \Delta(1700)} \frac{3}{2}^- \end{array}$    & \it 1656\er14  & 
            $\sqrt{M_\Delta^2 + \omega}$ & 1662 & 
            $\sqrt{12\lambda}$   & 1711 \\ \hline\hline

			$2$   & $0^+$ & $\mathbf{56}$ & $\frac{1}{2}$ & 
			\boldmath $N(1440) \frac{1}{2}^+$       & \it 1366\er 3  & 
            $\sqrt{M_\Delta^2 + 2\omega - A - \frac{\delta}{2}}$   & 1372  & 
            $\sqrt{8\lambda}$         & 1386 \\ \cline{4-10}

                  &       &               &  $\frac{3}{2}$ & 
            \boldmath $\Delta(1600) \frac{3}{2}^+$       & \it 1550\er 15  & 
            $\sqrt{M_\Delta^2 + 2\omega - A}$                      & 1570  & 
            $\sqrt{10\lambda}$                                     & 1574  \\ 
            \cline{2-10}

			      & $0^+$ & $\mathbf{70}$ & $\frac{1}{2}$ & 
            \boldmath $N(1710) \frac{1}{2}^+$                       & \it 1696\er10  & 
           $\sqrt{M_\Delta^2 + 2\omega - \frac{A}{2} - \frac{\delta}{4}}$   & 1724  & 
            $\sqrt{12\lambda}$                                               & 1711  \\ 
            \cline{4-10}

			      &       &               & $\frac{3}{2}$ & 
            $N'(1720) \frac{3}{2}^+$                    & \it 1725\er 30  & 
           $\sqrt{M_\Delta^2 + 2\omega - \frac{A}{2}}$           & 1801  & 
            $\sqrt{14\lambda}$                                    & 1848  \\ 
            \cline{4-10}

			      &       &               & $\frac{1}{2}$ & 
            $\Delta(1750) \frac{1}{2}^+$               & {\it 1770\er30}  & 
            $\sqrt{M_\Delta^2 + 2\omega - \frac{A}{2}}$           & 1801  & 
            $\sqrt{14\lambda}$                                    & 1848  \\ 
            \cline{2-10}

			      & $2^+$ & $\mathbf{56}$ & $\frac{1}{2}$ & 
            \boldmath \hspace{-1mm}$\begin{array}{l} {\displaystyle N(1720)} \frac{3}{2}^+ \\[-1mm] {\displaystyle N(1680)} \frac{5}{2}^+ \end{array}$    
                                                                     & \it 1689\er13  & 
            $\sqrt{M_\Delta^2 + 2\omega - \frac{2A}{5} - \frac{\delta}{2}}$   & 1696  & 
            $\sqrt{12\lambda}$                                                & 1711  \\ 
            \cline{4-10}

			      &       &               & $\frac{3}{2}$ & 
            \boldmath \hspace{-1mm}$\begin{array}{l} {\displaystyle \Delta(1910)} \frac{1}{2}^+ \\[-1mm] {\displaystyle \Delta(1920)} \frac{3}{2}^+ \\[-1mm] {\displaystyle \Delta(1905)} \frac{5}{2}^+ \\[-1mm] {\displaystyle \Delta(1950)} \frac{7}{2}^+ \end{array}$    
            &                                            \it 1859\er14   & 
            $\sqrt{M_\Delta^2 + 2\omega - \frac{2A}{5}}$         & 1850  & 
            $\sqrt{14\lambda}$                                   & 1848  \\ 
            \cline{2-10}

			      & $2^+$ & $\mathbf{70}$ & $\frac{1}{2}$ & 
            \hspace{-1mm}$\begin{array}{l} \mathbf{{\displaystyle N(1900)}} \frac{3}{2}^+ \\[-1mm] {\displaystyle N(1860)} \frac{5}{2}^+ \end{array}$    
            &                                                         \it 1891\er25  & 
            $\sqrt{M_\Delta^2 + 2\omega - \frac{A}{5} - \frac{\delta}{4}}$   & 1883  & 
            $\sqrt{14\lambda}$                                               & 1848  \\ 
            \cline{4-10}

			      &       &  \raisebox{5mm}{Q.M.}            & $\frac{3}{2}$ & 
            \hspace{-1mm}$\begin{array}{l} \mathbf{{\displaystyle N(1880)} \frac{1}{2}^+} \\[-1mm] {\displaystyle N(1965)} \frac{3}{2}^+ \\[-1mm] {\displaystyle N(2000)} \frac{5}{2}^+ \\[-1mm] {\displaystyle N(1990)} \frac{7}{2}^+ \end{array}$    
            &                                             \it 1978\er39  & 
            $\sqrt{M_\Delta^2 + 2\omega - \frac{A}{5}}$          & 1935  & 
            $\sqrt{16\lambda}$                                   & 1976 \\ 
            \cline{4-10}

			      &       &               & $\frac{1}{2}$ & 
            \hspace{-1mm}$\begin{array}{l}  \\[-1mm] {\displaystyle \Delta(2000)} \frac{5}{2}^+  \end{array}$    
            &                                             \it 2040\er80  & 
            $\sqrt{M_\Delta^2 + 2\omega - \frac{A}{5}}$          & 1935  & 
            $\sqrt{16\lambda}$                                   & 1976  \\ 
            \cline{2-10}

			      & $1^+$ & $\mathbf{20}$ & $\frac{1}{2}$ & 
            \hspace{-1mm}$\begin{array}{l} {\displaystyle N(2100)} \frac{1}{2}^+ \\[-1mm] {\displaystyle N(2040)} \frac{3}{2}^+ \end{array}$    
            &                       \it 2044\er 25  & 
            $\sqrt{M_\Delta^2 + 2\omega}$   & 2016  & 
                                            &   \\ \hline

		\end{tabular}
\vspace{-5mm}
\end{center}
\end{table}	
\begin{SCtable}[2][th]
\raisebox{-5mm}{
\caption{\label{AdS-nu}AdS/QCD values for $\nu$. 
}
\centering
\renewcommand{\arraystretch}{1.4}
\begin{tabular}{|l|cccccc|}
\hline
    && $^2N$  && $^4N$ && $\Delta$ \\
\hline
P\ =\ + &&    $\nu =L$      && $\nu =L+\frac12$&& $\nu =L+\frac12$\\
P\ =\ - && $\nu =L+\frac12$ && $\nu =L+1$ && $\nu =L+1$\\
\hline
\end{tabular}
\renewcommand{\arraystretch}{1.0}}
\end{SCtable}
\clearpage
The center-of-gravity masses 
$M_{\rm c.o.g.}$ in Table~\ref{baryons} are calculated as
\begin{eqnarray}
M_{\rm c.o.g.}=\frac{\sum_J J \cdot M_J}{\sum_J J},
\end{eqnarray}
where pole masses $M_J$ are used from Ref.~\cite{Sarantsev:2025lik}
or \cite{ParticleDataGroup:2024cfk}. The uncertainties are calculated by replacing $M_J$ 
by $(\delta M_J)^2$.

In this contribution, I give an interpretation of the ADS/QCD quantum numbers directly related
to physical effects: The nucleon mass is given by the zero point oscillation of the
excited string, the hyperfine structure splitting by instanton-induced interactions,
the quark masses are vanishingly small at the hadronic scale.

First we notice that $N^*$ resonances with an intrinsic quark spin $S=\frac32$
belonging to a spin quartet ($^4N$) and $\Delta^*$ resonances have (about) the same
mass, independent of of the quark-spin of $\Delta^*$ resonances. Expressed in
spin and isospin: $I=\frac12, S=\frac12$ have a lower mass than all other configurations.
If the spatial wave function is symmetric, the spin-isospin wave function 
is symmetric, too, and contains one component which is symmetric, and one which is
antisymmetric in spin and isospin. This latter part is influenced by instanton-induced
interactions. $^2N$ nucleons have a lower mass than $^4N$, $^2\Delta$
and $^4\Delta$ baryons due to instanton-induced interactions.

The classical rotation of a massless quark and a diquark can be described as a rotating
string of length $D$ with a constant mass distribution. With a mass in an interval 
$dr$ given by $\sigma dr$, the total
mass can be calculated to 
\begin{eqnarray}
M=\int_{-D/2} ^{+D/2} \frac{\sigma dr}{\sqrt{1-v^2}}=\frac{\pi\sigma D}{2} 
\label{string:M}
\end{eqnarray}
while the angular momentum is given by
\begin{eqnarray}
L= \int_{-D/2} ^{+D/2} \frac{\sigma v(r) r dr}{\sqrt{1-v^2}}=\frac{\pi\sigma D^2}{8}\,.   
\label{string:L}
\end{eqnarray}
Combining  these equations 
\begin{eqnarray}
M^2 = 2\pi\sigma L
\end{eqnarray}
In quantum mechanics, both oscillators have a zero-point energy $\pi\sigma$. Hence we write
\begin{eqnarray}
M^2 = 2\pi\sigma (L+\frac12+\frac12) = M_0 ^2 + a\cdot L\,.
\end{eqnarray}
For the $L=0$ ground state, we have $M_0 ^2=2\pi\sigma$, just the same as in AdS/QCD. The
mass of the proton can be understood as the zero-point energy of the two oscillators. 

Eq.~\ref{ads} quantifies the size of instanton-induced interactions
as $2\lambda$, and thus relates it to the Regge slope $4\lambda$ and
to the nucleon mass of $4\lambda$.  Mass generation, confinement, and chiral
symmetry breaking are all described by a single parameter. 

\begin{figure}[pt]
    \centering
    \includegraphics[width=0.8\linewidth]{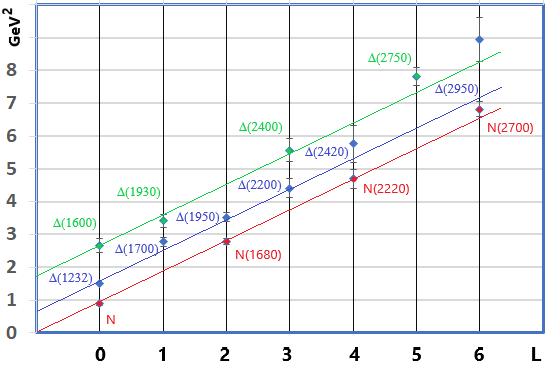}
    \caption{Squared masses (pole positions) of $N^*$ (red) and $\Delta^*$ (blue)
     resonances versus total angular momentum $L$. $\Delta^*$ resonances with radial
     excitation are shown in green.}
    \label{radials}
\end{figure}

The ADS/QCD approach has great achievements, not only in the calculation of mass
spectra of mesons and baryons but also in the derivation of space-like and time-like
formfactors. However, in some cases, the classification of confirmed baryons~
in Table~5.4 in Ref.~\cite{Brodsky:2014yha} is questionable. For instance, the 
3* $\Delta(1930)5/2^-$ assigned to a 70plet cannot have intrinsic $L=1,S=\frac32$
(but could exist with $L=2, S=1/2$ or $L=3, S=1/2$). However, it is more likely that
this state and the 3* $\Delta(1900)\frac12^-$ and the 2* $\Delta(1940)\frac32^-$ 
belong to a 56plet. The three $\Delta$ state must be accompanied by a nucleon doublet.
Indeed, there is a  spin doublet $N(1895)\frac12^-$ and $N(1875)\frac32^-$ while no $N\frac52^-$
has been reported. Probably, this is a required doublet. High-mass baryon resonances
often have $I=\frac12, S=\frac12$ or $I=\frac32, S=\frac32$, and their spin and orbital angular
momenta are aligned. Likely, $N(2700)\frac{11}{2}^-$ has intrinsic spin $S=\frac12$.

An interpretation of the baryon spectrum is given in Fig.~\ref{radials}. The Regge trajectory
is shown as a function of the intrinsic orbital angular momentum $L$, which can be
defined since spin-orbit interactions are small. Figure~\ref{radials} shows the squared
masses of $N^*$'s and $\Delta^*$ as in Fig.~\ref{regge}, but in addition
the squared masses of negative-parity $\Delta^*$ resonances and of radial excitations.

The only obstacle for AdS/QCD is $N(1710)\frac12^+$. In Ref.~\cite{Brodsky:2014yha}, it is the second
radial excitation of the nucleon and belongs to a 56plet. In quark models, it is
the companion of the Roper resonance in the 
$(70,0^+_2)$ multiplet with $S=1/2$. As a member of a 70-plet,  $N(1710)\frac{1}{2}^+$ has a 
wave function given by 
\begin{eqnarray}
 {\cal M_\mathcal{A}} = \phi_{00,11}\nonumber\,, \qquad
 {\cal M_\mathcal{S}} =\frac{1}{\sqrt{2}}( \phi_{10,00} - \phi_{01,00} ).    
\end{eqnarray} 
In $M_{\mathcal A}$ both, $\rho$ and $\lambda$, are excited to $l_\rho = l_\lambda=1$. In this case,
we expect that a sizable fraction of all decays proceed via an intermediate orbital excitation, that is, 
via $N(1535)\frac{1}{2}^-\pi$, $N(1520)\frac{3}{2}^-\pi$, or $N\sigma$. In the Review of Particle Physics, the branching ratio
to $N(1535)\frac{1}{2}^-\pi$ is listed with (9-21)\%, for $N\sigma$ an upper limit of 16\% is quoted.
These decay modes favor the quark model
interpretation of $N(1710)\frac{1}{2}^+$ as member of a 70plet. This interpretation
is supported in quark-model calculations. 

The Table~\ref{AdS-nu} quotes values of $\nu$ for positive-parity baryons in a SU(6) 56plet
and negative-parity baryons in a SU(6) 70plet. There are, however, baryons that do not fall 
into these two categories, $N(1710)\frac{1}{2}^+$ is one example. $\Delta(1930)\frac52^-$ with
$L=1,S=\frac32$ belonging to a 56plet is a second example. However, with a small change,
one can calculate masses of these additional states as well, see Table~\ref{AdS-nu2}.
In the harmonic oscillator approximation, $N(1710)\frac{1}{2}^+$ contains with
50\% probability $n=1, L=0$, but with the same probability ($n_\rho+n_\lambda=0, l_\rho+\l_\lambda=2$). The former yields 
an excitation of $10\lambda$, the latter of $14\lambda$, with $12\lambda$ as mean value. $\Delta(1750)\frac12^+$
is then expected at $14\lambda$. In other cases, the wave functions are more complicated, and
even fractional values for $\nu$ can result.

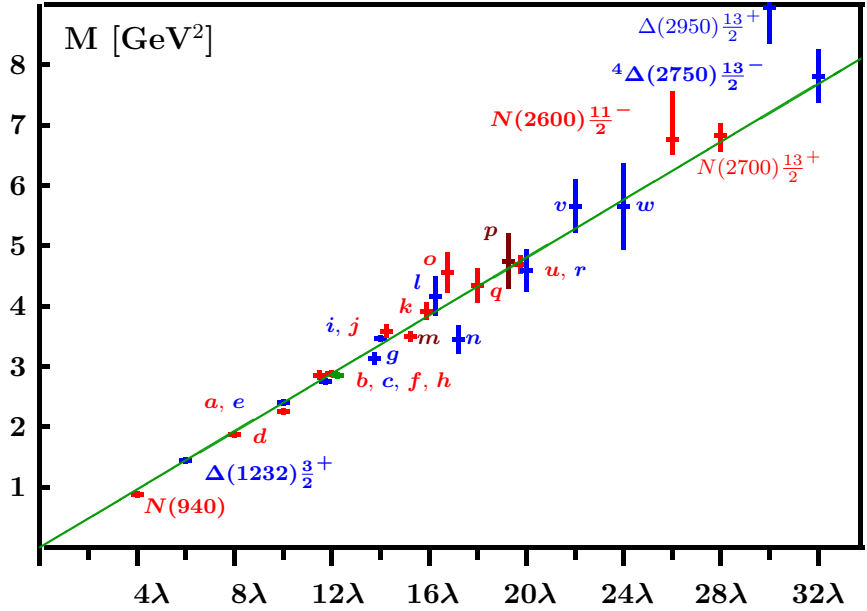
\begin{figure}[thb]
\begin{center}
\setlength{\unitlength}{0.8mm}
\hspace*{-4mm}
\linethickness{0.6mm}
\begin{picture}(150.00,90.00)
\put(9.00,00.00){ \line(1,0){135.00}}
\put(10.00,0.00){\line(0,1){90.00}}
\put(9.00,90.00){ \line(1,0){135.00}}
\put(145.00,0.00){\line(0,1){90.00}}
\put(10.00,-3.00){\line(0,1){3.00}}
\put(18.00,-2.00){\line(0,1){2.00}}
\put(26.00,-3.00){\line(0,1){3.00}}
\put(34.00,-2.00){\line(0,1){2.00}}
\put(42.00,-3.00){\line(0,1){3.00}}
\put(50.00,-2.00){\line(0,1){2.00}}
\put(58.00,-3.00){\line(0,1){3.00}}
\put(66.00,-2.00){\line(0,1){2.00}}
\put(74.00,-3.00){\line(0,1){3.00}}
\put(82.00,-2.00){\line(0,1){2.00}}
\put(90.00,-3.00){\line(0,1){3.00}}
\put(98.00,-2.00){\line(0,1){2.00}}
\put(106.00,-3.00){\line(0,1){3.00}}
\put(114.00,-2.00){\line(0,1){2.00}}
\put(122.00,-3.00){\line(0,1){3.00}}
\put(130.00,-2.00){\line(0,1){2.00}}
\put(138.00,-3.00){\line(0,1){3.00}}
\put(8.00,10.00){ \line(1,0){2.0}}
\put(8.00,20.00){ \line(1,0){2.0}}
\put(8.00,30.00){ \line(1,0){2.0}}
\put(8.00,40.00){ \line(1,0){2.0}}
\put(8.00,50.00){ \line(1,0){2.0}}
\put(8.00,60.00){ \line(1,0){2.0}}
\put(8.00,70.00){ \line(1,0){2.0}}
\put(8.00,80.00){ \line(1,0){2.0}}
\put(5.00,7.50){\makebox(10.00,5.00)[l]{\bf 1}}
\put(5.00,17.50){\makebox(10.00,5.00)[l]{\bf 2}}
\put(5.00,27.50){\makebox(10.00,5.00)[l]{\bf 3}}
\put(5.00,37.50){\makebox(10.00,5.00)[l]{\bf 4}}
\put(5.00,47.50){\makebox(10.00,5.00)[l]{\bf 5}}
\put(5.00,57.50){\makebox(10.00,5.00)[l]{\bf 6}}
\put(5.00,67.50){\makebox(10.00,5.00)[l]{\bf 7}}
\put(5.00,77.50){\makebox(10.00,5.00)[l]{\bf 8}}

\put(25.00,8.80){\rd\line(1,0){2.00}}
\put(26.00,8.30){\rd\line(0,1){1.00}}
\put(27.00,4.00){\makebox(15.00,5.00)[l]{\boldmath\scriptsize\rd $N(940)$ }}

\put(33.00,14.40){\bl\line(1,0){2.00}}                         
\put(34.00,13.90){\bl\line(0,1){1.00}}
\put(37.00,10.00){\makebox(15.00,5.00)[l]
{\boldmath\scriptsize\bl $\Delta(1232)\frac32^+$ }}

\put(41.00,18.66){\rd\line(1,0){2.00}} 
\put(42.00,18.16){\rd\line(0,1){1.00}}
\put(45.00,16.00){\makebox(15.00,5.00)[l]{\boldmath\scriptsize\rd $d$ }}

\put(49.00,22.59){\rd\line(1,0){2.00}}
\put(50.00,22.09){\rd\line(0,1){1.00}}

\put(49.00,24.03){\bl\line(1,0){2.00}} 
\put(50.00,23.53){\bl\line(0,1){1.00}}

\put(37.00,21.00){\makebox(50.00,5.00)[l]{\boldmath\scriptsize\rd $a$, \bl $e$}}

\put(55.0,28.56){\rd\line(1,0){2.00}} 
\put(56.0,27.75){\rd\line(0,1){1.62}}

\put(56.0,27.42){\bl\line(1,0){2.00}}
\put(57.0,26.92){\bl\line(0,1){1.00}}

\put(57.0,28.76){\rd\line(1,0){2.00}} 
\put(58.0,28.26){\rd\line(0,1){1.00}}

\put(58.0,28.52){\gr\line(1,0){2.00}} 
\put(59.0,28.02){\gr\line(0,1){1.00}}

\put(62.00,25.00){\makebox(35.00,5.00)[l]{\boldmath\scriptsize \rd $b$, \bl $c$, \rd $f$, \rd $h$ }}

\put(64.00,31.33){\bl\line(1,0){2.00}} 
\put(65.00,30.27){\bl\line(0,1){2.12}}

\put(65.00,34.56){\bl\line(1,0){2.00}} 
\put(66.00,34.06){\bl\line(0,1){1.00}}

\put(66.00,35.76){\rd\line(1,0){2.00}} 
\put(67.00,34.65){\rd\line(0,1){2.27}}
 
\put(67.00,29.00){\makebox(15.00,5.00)[l]{\boldmath\scriptsize  \bl $g$ }}
\put(57.00,34.00){\makebox(15.00,5.00)[l]{\boldmath\scriptsize \bl $i$, \rd $j$ }}

\put(70,34.89){\rd\line(1,0){2.00}} 
\put(71.00,34.03){\rd\line(0,1){1.72}}

\put(72.00,32.00){\makebox(15.00,5.00)[l]{\boldmath\scriptsize \br $m$ }}

\put(72.50,39.12){\rd\line(1,0){2.00}} 
\put(73.50,37.70){\rd\line(0,1){2.84}}
\put(69.00,37.50){\makebox(15.00,5.00)[l]{\boldmath\scriptsize\rd $k$ }}
\put(74.00,41.62){\bl\line(1,0){2.00}} 
\put(75.00,38.36){\bl\line(0,1){6.53}}
\put(71.50,41.50){\makebox(15.00,5.00)[l]{\boldmath\scriptsize\bl $l$ }}

\put(76.00,45.58){\rd\line(1,0){2.00}} 
\put(77.00,42.21){\rd\line(0,1){6.75}}
\put(73.00,45.00){\makebox(15.00,5.00)[l]{\boldmath\scriptsize \rd $o$ }}

\put(77.80,34.52){\bl\line(1,0){2.00}} 
\put(78.80,32.18){\bl\line(0,1){4.68}}
\put(80.00,32.0){\makebox(15.00,5.00)[l]{\boldmath\scriptsize \bl $n$ }}

\put(81.00,43.47){\rd\line(1,0){2.00}} 
\put(82.00,40.63){\rd\line(0,1){5.67}}
\put(84.00,39.50){\makebox(15.00,5.00)[l]{\boldmath\scriptsize\rd $q$ }}

\put(86.00,47.44){\br\line(1,0){2.00}} 
\put(87.00,42.82){\br\line(0,1){9.23}}
\put(83.00,49.50){\makebox(15.00,5.00)[l]{\boldmath\scriptsize\br $p$ }}

\put(88.00,46.87){\rd\line(1,0){2.00}} 
\put(89.00,45.36){\rd\line(0,1){3.03}}

\put(89.00,45.88){\bl\line(1,0){2.00}} 
\put(90.00,42.32){\bl\line(0,1){7.10}}

\put(93.00,42.90){\makebox(15.00,5.00)[l]{\boldmath\scriptsize\rd $u$, \bl  $r$ }}

\put(97.00,56.50){\bl\line(1,0){2.00}}
\put(98.00,52.03){\bl\line(0,1){8.94}}
\put(83.00,54.00){\makebox(15.00,5.00)[r]{\boldmath\scriptsize\bl $v$ }}
\put(105.00,56.50){\bl\line(1,0){2.00}}
\put(106.00,49.37){\bl\line(0,1){14.26}}
\put(108.00,54.00){\makebox(15.00,5.00)[l]{\boldmath\scriptsize\bl $w$ }}
\put(121.00,68.23){\rd\line(1,0){2.00}} 
\put(122.00,65.53){\rd\line(0,1){4.70}} 
\put(113.00,67.60){\rd\line(1,0){2.00}} 
\put(114.0,65.00){\rd\line(0,1){10.60}}
\put(84.00,68.60){\makebox(15.00,5.00)[l]{\boldmath\scriptsize\rd $N(2600)\frac{11}{2}^-$ }}
\put(118.00,60.60){\makebox(15.00,5.00)[l]{\boldmath\scriptsize\rd  \unboldmath  $N(2700)\frac{13}{2}^+$ }}
\put(137.00,78.06){\bl\line(1,0){2.00}} 
\put(138.00,73.59){\bl\line(0,1){8.94}}
\put(104.00,76.00){\makebox(15.00,5.00)[l]{\boldmath\scriptsize\bl $^4\Delta(2750)\frac{13}{2}^-$ }}
\put(129.00,89.40){\bl\line(1,0){2.00}} 
\put(130.00,83.42){\bl\line(0,1){6.90}}
\put(115.00,84.00){\makebox(15.00,5.00)[r]{\bl\scriptsize $\Delta(2950)\frac{13}{2}^+$ }}
\put(10.00,-0.10){\bf\gr\line(5,3){135.00}}
\put(10.00,0.00){\bf\gr\line(5,3){135.00}}
\put(10.00,0.10){\bf\gr\line(5,3){135.00}}
\put(14.00,83.00){\makebox(10.00,2.00)[l]{\bf\large M [GeV$^2$]}}
\put(10.00,-10.00){\makebox(15.00,5.00)[r]{\bf   }}
\put(18.00,-10.00){\makebox(15.00,5.00)[r]{\boldmath $4\lambda$ }}
\put(26.00,-10.00){\makebox(15.00,5.00)[r]{\bf   }}
\put(34.00,-10.00){\makebox(15.00,5.00)[r]{\boldmath $8\lambda$ }}
\put(42.00,-10.00){\makebox(15.00,5.00)[r]{\bf  }}
\put(50.00,-10.00){\makebox(15.00,5.00)[r]{\boldmath $12\lambda$ }}
\put(58.00,-10.00){\makebox(15.00,5.00)[r]{\bf  }}
\put(64.00,-10.00){\makebox(15.00,5.00)[r]{\boldmath $16\lambda$}}
\put(74.00,-10.00){\makebox(15.00,5.00)[r]{\bf }}
\put(80.00,-10.00){\makebox(15.00,5.00)[r]{\boldmath  $20\lambda$}}
\put(90.00,-10.00){\makebox(15.00,5.00)[r]{\bf }}
\put(96.00,-10.00){\makebox(15.00,5.00)[r]{\boldmath  $24\lambda$}}
\put(106.00,-10.00){\makebox(15.00,5.00)[r]{\bf }}
\put(113.00,-10.00){\makebox(15.00,5.00)[r]{\boldmath $2\boldmath8\lambda$ }}
\put(122.00,-10.00){\makebox(15.00,5.00)[r]{\bf  }}
\put(129.00,-10.00){\makebox(15.00,5.00)[r]{\boldmath $32\lambda$ }}
\put(138.00,-10.00){\makebox(15.00,5.00)[r]{\bf  }}
\end{picture}
\vspace{4mm}
\end{center}
\caption{\label{fig:Regge-AdS} The squared mass of $N^*$'s and $\Delta^*$'s as a function
of the masses predicted in AdS/QCD. For multiplets, the center-of-gravity mass is shown. \\
\begin{tiny}\noindent
a:  $N(1535)\frac12^-$, $ N(1520)\frac32^-$; b: $N(1650)\frac12^-$, $ N(1700)\frac32^-$, $ N(1675)\frac52^-$;
c: $\Delta(1620)\frac12^-$, $\Delta(1700)\frac32^-$,\\[-1ex] 
h: $N(1720)\frac32^+$, $N(1680)\frac52^+$; d:  $N(1440)\frac12^+$; e: $\Delta(1600)\frac32^+$; f: 
$N(1710)\frac12^+$; g: $\Delta(1750)\frac12^+$;\\[-1ex]
h:~$N(1720)\frac32^+$, $N(1680)\frac52^+$; i:~$\Delta(1910)\frac12^+$, $\Delta(1930)\frac32^+$, $\Delta(1905)\frac52^+$, $\Delta(1950)\frac72^+$;  j:~$N(1900)\frac32^+$, $ N(1860)\frac52^-$; \\[-1ex]  
k: $N(1880)\frac12^+$, $ N(1965)\frac32^+$, $ N(2000)\frac52^+$, $N(1990)\frac72^+$;
l: $\Delta(2000)\frac52^+$; m: $N(1895)\frac12^-$, $ N(1875)\frac32^-$; \\[-1ex]
n: $\Delta(1900)\frac12^-$, $\Delta(1940)\frac32^-$, $\Delta(1930)\frac52^-$:
o: $N(2150)\frac{1}{2}^-$, $N(2120)\frac{3}{2}^-$; p: $\Delta(2150)\frac{1}{2}^-$, $\Delta(2190)\frac{3}{2}^-$; \\[-1ex]
q:~$N(2060)\frac52^-$, $N(2190)\frac72^-$; r: $\Delta(2210)\frac52^-$, $\Delta(2200)\frac72^-$; 
s: $ N(2250)\frac92^-$; u: $N(2220)\frac92^+$; \\[-1ex] 
v: $\Delta(2390)\frac{7}{2}^+$, $\Delta(2300)\frac{9}{2}^+$,$\Delta(2420)\frac{11}{2}^+$;
w: $\Delta(2350)\frac{5}{2}^-$, $\Delta(2400)\frac{9}{2}^-$.
\end{tiny}
}
\end{figure}

Figure~\ref{fig:Regge-AdS} shows the squared masses of (nearly) all resonances as a function of $\lambda$.
Most resonances are at least compatible with the straight line.

\begin{SCtable}[2][th]
\caption{AdS/QCD values for $\nu$. }
\centering
\renewcommand{\arraystretch}{1.4}
  \setlength{\tabcolsep}{5pt}
\begin{tabular}{cc|ccccc}
\hline 
  $D^P$  && $^2N$  && $^4N$ && $\Delta$ \\
\hline
56$^+$&&  $\nu =l$         && $\nu =l+\frac12$&& $\nu =l+\frac12$\\
else&&  $\nu =l+\frac12$ && $\nu =l+1$      && $\nu =l+1$\\
\hline 
\end{tabular}
\renewcommand{\arraystretch}{1.0}
\label{AdS-nu2}
\end{SCtable} 

\section{Quark model approach}

Quark models describe the spectrum of baryon resonances as excitations in the dynamics
of three constituent quarks. Since the early model of Isgur and Karl~\cite{Isgur:1977ef,Isgur:1978xj,Isgur:1978wd}, relativistic
corrections have been implemented~\cite{Capstick:1986ter} or fully relativistic calculations were 
carried out~\cite{Loring:2001kx,Loring:2001ky,Loring:2001kv}.
All models agree on linear confining potential, but a dispute arose on the residual 
quark-quark interaction. Can these be treated as an effective one-gluon exchange~\cite{Isgur:1977ef,Isgur:1978xj,Isgur:1978wd,Capstick:1986ter} 
or by exchanges of pseudoscalar mesons between quarks~\cite{Glozman:1997ag,Glozman:1997fs}? 
Are instanton-induced interactions
at work~\cite{Loring:2001kx,Loring:2001ky,Loring:2001kv}? 

In quark models, the three-body system is reduced to two oscillations
and a trivial center-of-mass motion. The two oscillators in the variables $\rho$ and $\lambda$ represent the oscillation of a diquark relative to the third quark ($\lambda$)
and the oscillation within the diquark. Symmetrization guarantees that all quarks are
treated equally. Both oscillators can be excited radially ($n_\rho, n_\lambda$) or carry angular momenta ($l_\rho, l_\lambda$). We call $n=n_\rho + n_\lambda$ radial
and $l=l_\rho + l_\lambda$ orbital excitation quantum number. The orbital angular momentum
$\vec L=\vec l_\rho + \vec l_\lambda$ combines with the total quark spin $\vec S$ to the
particle spin $\vec J=\vec L+\vec S$. For $S=1/2$, two resonances with $J=\pm\frac12$ are
expected, for $S=3/2$, a quartet develops with $J=-\frac32, \cdots , J=+\frac32$. In the absence
of spin-orbit interactions, the spin-doublet and the spin-quartet would be degenerate in mass.

\begin{table}[pt]
\caption{\label{QM} Mass pattern of baryon resonances in the non-relativistic quark model with
a linear confinement potential.\\[-2ex] }
\begin{tabular}{cc}
\raisebox{8mm}{
\begin{minipage}[d]{0.40\textwidth}
    \centering
\renewcommand{\arraystretch}{1.4}    
    \begin{tabular}{lc}
    \hline\hline
 $(D,J^P)_N$   & lin. conf.   \\  \hline
 $(56,0^+)_0$ & $M_0$        \\
 $(70,1^-)_1$  & $M_0+\omega$  \\
 $(56,0^+)_2$ & $M_0+2\omega-A$\\
 $(70,0^+)_2$ & $M_0+\omega-\frac12 A$ \\
 $(56,2^+)_2$ & $M_0+2\omega-\frac25 A$\\
 $(70,2^+)_2$ & $M_0+2\omega-\frac15 A$\\
 $(20,1^+)_2$ & $M_0+2\omega$  \\
 \hline\hline
    \end{tabular}
\renewcommand{\arraystretch}{1.0}    
\end{minipage}
}
&
\begin{minipage}[d]{0.50\textwidth}
\renewcommand{\arraystretch}{1.4}    
\begin{tabular}{ll}
\multicolumn{2}{l}{\footnotesize Table 3. Fraction of ``good diquarks'' in  the}\\[-1ex]
\multicolumn{2}{l}{\footnotesize wave function of $N^*$ resonances. For the nuc-}\\[-1ex] 
\multicolumn{2}{l}{\footnotesize leon, the fraction is $\frac12$. It is responsible for the } \\[-1ex]
\multicolumn{2}{l}{\footnotesize  $\Delta -N$ mass splitting.} \\[1ex]
\hline\hline
$\rm I_{sym}  = \frac12$& for S=$\frac12$ $N^*$'s in a 56plet; \\
$\rm I_{sym}  = \frac14$& for S=$\frac12$  $N^*$'s  in a 70plet;\\
$\rm I_{sym}  =   0$& for S=$\frac32$   $N^*$'s \\
$\rm I_{sym}  =   0$&   for   $N^*$'s  in a 20plet;\\
$\rm I_{sym}  =   0$&   for  $\Delta^*$'s.  \\
\hline\hline
&\\
&\\
&\\
\end{tabular}
\renewcommand{\arraystretch}{1.0}    
\end{minipage}
\end{tabular}
\vspace{-22mm}
\end{table}
The quark-quark and quark-diquark potential is linearly increasing, but quark models
start with the harmonic-oscillator approximation. Its eigenvalues increase linearly
as $M_0+ (l+2n)\omega$. The linear potential is taken into
account  perturbatively yielding corrections to the eigenvalues for the lowest excitation levels listed in
Table~\ref{QM}.
As mentioned above, instanton-induced interaction lower the mass of resonances due to the fraction
in their wave function that is antisymmetric in spin and isospin (or spin and flavor). 
This fraction is shown in Table 3.

In Table~\ref{baryons}, low-mass baryon resonances are listed with their SU(6) classification.
The labels $L, S$ and $n$ refer to the internal orbital angular momentum, internal spin and 
radial excitation quantum number. The mass values suggest that spin-orbit interaction must be 
small. In the third row we therefore give the mean (pole) mass of a multiplet, the error
covers the range of masses within the multiplet. It is calculated from the maximum spread of results
divided by $\sqrt 12$.

In the fourth row we give formulae and results for the masses of baryons in a multiplet. The
potential parameter $A$, the string constant $a$, and the variable defining instanton-induced
interactions $b$ are chosen as $A=1.6$, $a=1.3$, $b=1.16$, all in GeV$^2$ units. The similarity of $a$ 
and $b$ reminds of the common origin of Regge slope and hyperfine structure. In consideration
of the simplicity of the formulas, the agreement between data and fit is surprising. 

\section{Conclusions}
We have seen that the gross features of the spectrum of light-baryon resonances find simple explanations. 
Baryon with angular momentum excitation can be understood in a very simple picture of a rotating string. AdS/QCD
provides a more detailed and preciser view of the spectrum. Its success underlines the deep
connection between the mass of baryons, the string tension and the hyperfine interaction.
Surprisingly, the non-relativistic quark model provides for a detailed view of the full 
richness of the three-body dynamics.


\begin{thebibliography}{99}

\bibitem{Diakonov:2002mb}
D.~Diakonov,
``Instantons and baryon dynamics,''
9th International Conference on the Structure of Baryons,
Newport News, VA, United States, 3-8 March 2002.

\bibitem{Diakonov:1987ty}
D.~Diakonov, V.~Y.~Petrov and P.~V.~Pobylitsa,
``A Chiral Theory of Nucleons,''
Nucl. Phys. B \textbf{306}, 809 (1988).

\bibitem{Diakonov:1995qy}
D.~Diakonov, M.~V.~Polyakov and C.~Weiss,
``Hadronic matrix elements of gluon operators in the instanton vacuum,''
Nucl. Phys. B \textbf{461}, 539-580 (1996).

\bibitem{Klempt:2002vp}
E.~Klempt,
``A Mass formula for baryon resonances,''
Phys. Rev. C \textbf{66}, 058201 (2002)

\bibitem{Klempt:2010du}
E.~Klempt,
``Nucleon Excitations,''
Chin. Phys. C \textbf{34}, no.9, 1241-1246 (2010).

\bibitem{Klempt:2002cu}
E.~Klempt,
``Baryon resonances and strong QCD,''
[arXiv:nucl-ex/0203002 [nucl-ex]].

\bibitem{Diakonov:2004jn}
D.~Diakonov, N.~Gromov, V.~Petrov and S.~Slizovskiy,
``Quantum weights of dyons and of instantons with nontrivial holonomy,''
Phys. Rev. D \textbf{70}, 036003 (2004).

\bibitem{Diakonov:2007nv}
D.~Diakonov and V.~Petrov,
``Confining ensemble of dyons,''
Phys. Rev. D \textbf{76}, 056001 (2007).

\bibitem{Brodsky:2014yha}
S.~J.~Brodsky, G.~F.~de Teramond, H.~G.~Dosch and J.~Erlich,
``Light-Front Holographic QCD and Emerging Confinement,''
Phys. Rept. \textbf{584}, 1-105 (2015).

\bibitem{Sarantsev:2025lik}
A.~V.~Sarantsev {\it et al.},
``Decays of $N^*$ and $\Delta^*$ resonances into $N\rho$, $\Delta\pi$, and $N\sigma$,''
[arXiv:2503.16636 [nucl-th]].

\bibitem{ParticleDataGroup:2024cfk}
S.~Navas \textit{et al.} [Particle Data Group],
``Review of particle physics,''
Phys. Rev. D \textbf{110}, no.3, 030001 (2024).

\bibitem{Isgur:1977ef}
N.~Isgur and G.~Karl,
``Hyperfine Interactions in Negative Parity Baryons,''
Phys. Lett. B \textbf{72}, 109 (1977).

\bibitem{Isgur:1978xj}
N.~Isgur and G.~Karl,
``P Wave Baryons in the Quark Model,''
Phys. Rev. D \textbf{18}, 4187 (1978).

\bibitem{Isgur:1978wd}
N.~Isgur and G.~Karl,
``Positive Parity Excited Baryons in a Quark Model with Hyperfine Interactions,''
Phys. Rev. D \textbf{19}, 2653 (1979)
[erratum: Phys. Rev. D \textbf{23}, 817 (1981)].

\bibitem{Capstick:1986ter}
S.~Capstick and N.~Isgur,
``Baryons in a relativized quark model with chromodynamics,''
Phys. Rev. D \textbf{34}, no.9, 2809-2835 (1986).

\bibitem{Loring:2001kx}
U.~Loring, B.~C.~Metsch and H.~R.~Petry,
``The Light baryon spectrum in a relativistic quark model with instanton induced quark forces: The Strange baryon spectrum,''
Eur. Phys. J. A \textbf{10}, 447-486 (2001).

\bibitem{Loring:2001ky}
U.~Loring, B.~C.~Metsch and H.~R.~Petry,
``The Light baryon spectrum in a relativistic quark model with instanton induced quark forces: The Strange baryon spectrum,''
Eur. Phys. J. A \textbf{10}, 447-486 (2001)

\bibitem{Loring:2001kv}
U.~Loring, K.~Kretzschmar, B.~C.~Metsch and H.~R.~Petry,
``Relativistic quark models of baryons with instantaneous forces: Theoretical background,''
Eur. Phys. J. A \textbf{10}, 309-346 (2001).

\bibitem{Glozman:1997ag}
L.~Y.~Glozman, W.~Plessas, K.~Varga and R.~F.~Wagenbrunn,
``Unified description of light and strange baryon spectra,''
Phys. Rev. D \textbf{58}, 094030 (1998).

\bibitem{Glozman:1997fs}
L.~Y.~Glozman, Z.~Papp, W.~Plessas, K.~Varga and R.~F.~Wagenbrunn,
``Effective Q-Q interactions in constituent quark models,''
Phys. Rev. C \textbf{57}, 3406-3413 (1998).
\end{thebibliography}
\end{document}